# Flow in the Driven Cavity Calculated by the Lattice Boltzmann Method

W. Miller

*CRS4 - Centre for Advanced Studies, Research and Developement, Via N.Sauro 10, I-90123 Cagliari, Italy*

The lattice Boltzmann method with enhanced collisions and rest particles is used to calculate the flow in a two-dimensional lid-driven cavity. The abilitity of this method to compute the velocity and the pressure of an incompressible fluid in a geometry with Dirichlet and Neumann boundary conditions is verified by calculating a test-problem where the analytical solution is known. Different parameter configurations have been tested for Reynolds numbers from $Re = 10$ to $Re = 2000$. The vortex structure for a more generalized lid-driven cavity problem with a non-uniform top speed has been studied for various acpect ratios.

## I. INTRODUCTION

Lattice gas automata (LGA) are a rather new technique in the world of fluid mechanics, but since the first discovery of a LGA by Frisch, Hasslacher and Pomeau [1], which reproduces all terms of the Navier-Stokes equations, they have undertaken a fast development. With the introduction of the lattice Boltzmann method (LBM) by McNamara and Zanetti [2] the statistical noise, present in the LGA, has been removed. Higuera, Succi and Benzi developed the LBM with 'enhanced collisions', where the foremost complicated collision operator was replaced by a much simpler one [3]. A further simplification can be achieved if the collision term is described by a simple relaxation. The LBM becomes then a BGK model Qian [4]. The introduction of rest particles by H.Chen, S.Chen and Matthaeus [5] removed the unphysical factor in the pressure term of the previous models and allowed the correct calculation of the pressure.

In this paper the LBM with enhanced collisions and with rest particles is used to study the flow and pressure distribution in a lid-driven cavity. There exists a particular benchmark problem introduced by Shih, Tan, and Hwang [6], where the analytical solution is known. This gives the possibility to test the numerical code for both Dirichlet and Neumann boundary conditions. Having verified that this method reproduces the analytic solution with small errors comparable to other numerical methods I go beyond the theoretical solvable problem and change the aspect ratio of the system. I study the change of the vortex structure during the increase of the aspect ratio for a small and a large Reynolds number ($Re = 50$ and 1000). The occurring inflexional shear between the flows of opposite direction remains stable for $Re = 50$ while for $Re = 1000$ it is unstable.

## II. THE LATTICE BOLTZMANN MODEL

In general a lattice gas automaton consists of particles moving on links ($i = 1, .., M_{\vec{c}}$) of a lattice from one node $\vec{r}_*$ to the next, where they can collide with one another. The time evolution for the population $n_i$ for particles with speed $\vec{c_i}$ is then given by a moving step and a collision, described by the operator $\Omega_{ij}$:[1]

$$n_i(\vec{r}_* + \vec{c}_i, t_* + 1) = n_i(\vec{r}_*, t_*) + \Omega_{ij} \qquad (1)$$

Both operations, i.e. move and collision, must conserve mass and momentum. If the mean population are denoted with $N_i(\vec{r}_*, t_*) \equiv < n_i(\vec{r}_*, t_*) >$ the density $\rho$ and velocity $\vec{u}$ is given by $\rho = \sum_i N_i$ and $\vec{u} = \frac{1}{\rho}\sum_i N_i\vec{c}_i$, respectively.

In the continuum limit the model should converge towards the Navier-Stokes equations. It has been shown that in two dimensions (2d) the hexagonal lattice is a proper choice [1], and in three dimensions (3d) it is the face-centered hypercubic (FCHC) lattice [7]. Assuming that all particles have the same speed and obey Fermi statistics[2] the equilibrium distribution for the population $N_i$ is given by:

$$N_i^{eq} = \left[1 + e^{h + \vec{q}\vec{c}_i}\right]^{-1} \qquad (2)$$

where $h$ and $\vec{q}$ are Langragian multipliers associated with the conserved quantities mass and momentum. For small Mach numbers one can compute $h$ and $\vec{q}$ pertubatively in $\vec{u}$ and end up with the following expansion for $N_i^{eq}$:

$$N_i^{eq} = d + \underbrace{d\frac{D}{c^2}(c_i)_\alpha u_\alpha}_{N_i^{eq,1}} + \underbrace{d\frac{D(D+2)}{2c^4} g(\rho) Q_{i\alpha\beta} u_\alpha u_\beta}_{N_i^{eq,2}} + \mathcal{O}(u^3) \qquad (3)$$

where

$$c = |\vec{c}_i| \quad , \quad g(\rho) = \frac{D}{(D+2)}\frac{1-2d}{1-d} \quad ,$$

$$\text{and} \quad Q_{i\alpha\beta} = (c_i)_\alpha (c_i)_\beta - \frac{c^2}{D}\delta_{\alpha\beta}$$

$d$ is the density per link $d = \frac{\rho}{M_{\vec{c}}}$.

The continuum limit is performed as a multiscale expansion in a small parameter $\varepsilon$, which can be identified with the local Knudsen number. The actual population $N_i$ is expanded around its equilibrium:

$$N_i = N_i^{eq} + \varepsilon N_i^{neq,1} + \mathcal{O}(\varepsilon^2)$$

---

[1] The time step is set equal to 1.

[2] The Fermi statistic has its origin in the lattice gas models with Boolean variables. Within the lattice Boltzmann approach it is possible to choose a different distribution function, see e.g. [4].

Using the mass and momentum conservation for the dynamics of the lattice automaton one arrives at the Navier-Stokes equations [1]:

$$\frac{\partial \rho}{\partial t} + \frac{\partial u_\beta}{\partial r_\beta} = 0 \qquad (4a)$$

$$\frac{\partial}{\partial t}(\rho u_\alpha) + \frac{\partial \Pi_{\alpha\beta}}{\partial r_\beta} = \frac{\partial S_{\alpha\beta}}{\partial r_\beta} + \mathcal{O}(\varepsilon^3 u) + \mathcal{O}(\varepsilon^2 u^2) + \mathcal{O}(\varepsilon u^3) \qquad (4b)$$

$t = \varepsilon t_* + \varepsilon^2 t_*$ and $\vec{r} = \varepsilon \vec{r}_*$ are the continous time and space varibiable, respectively. The viscous stress tensor $S_{\alpha\beta}$ is given by:

$$S_{\alpha\beta} = \nu \left( \frac{\partial(\rho u_\beta)}{\partial r_\alpha} + \frac{\partial(\rho u_\alpha)}{\partial r_\beta} - \frac{2}{D}\delta_{\alpha\beta}\frac{\partial(\rho u_\gamma)}{\partial r_\gamma} \right) \qquad (5)$$

$\nu$ is the viscosity of the system, which depends on the details of the collisions. An expression for $\nu$ will be given below. The momentum flux tensor $\Pi_{\alpha\beta}$ has the following form:

$$\Pi_{\alpha\beta} = \underbrace{\rho c_s^2 \left(1 - g(\rho)\frac{u^2}{c^2}\right)}_{p} + \rho g(\rho) u_\alpha u_\beta \qquad (6)$$

$c_s = \sqrt{(c^2/D)}$ is the sound speed and $p$ the pressure. One disadvantage of the lattice automaton with Boolean variables is the need of performing a statistical average numerically. This step can be made obsolete by using the mean population $N_i$, which defines the lattice Boltzmann model (LBM). If the Boltzmann approximation is taken for the collision, the collision operator $\Omega_{ij}$ can be approximated by an expression linearized in the non-equilibrium part $N_i^{neq} = N_i - N_i^{eq}$ of the population numbers:

$$\Omega_{ij} \approx \mathcal{A}_{ij}(N_j - N_j^{eq}) \qquad (7)$$

The elements $a_{ij}$ of the matrix $\mathcal{A}_{ij}$ describe the scattering between direction $i$ and $j$ and depend on the angle between the directions only. This approach was introduced by Higuera, Succi and Benzi and is called LBM with *enhanced collisions*. Because $\mathcal{A}_{ij}$ is cyclic and symmetric, their elements $a_{ij}$ can be expressed in terms of the three non-zero eigenvalues of the matrix: $\lambda$, $\sigma$ and $\tau$ [3]. The influence of $\sigma$ and $\tau$ on the results is negligible, and they both can be set equal to -1 [3]. The other eigenvalue is related to the viscosity via:

$$\nu = -\frac{1}{3}\left(\frac{1}{\lambda} + \frac{1}{2}\right) \qquad (8)$$

Furthermore, it can be shown that $-\frac{1}{\lambda}$ is the relaxation time, in which the population $N_i$ at a node converges to its equilibrium $N_i^{eq}$ [8].
For the one-speed model, described above, the pressure $p$ depends explicitely on the velocity $u$, as it can be seen from equation (6). This unsatisfying feature can be removed by introducing a reservoir of rest particles with a density $d_0$ which is a function of the density $d$ [5]. The total density of the system is then given by:

$$\rho = d_0 + M_{\vec{c}} d \tag{9}$$

In the following I will restrict myself to the FCHC lattice where $D = 4$, $M_{\vec{c}} = 24$ and $c = 2$. In this model choose the density of the rest particles is chosen as:

$$d_0 = \left( \frac{3}{2} \frac{(1-d)}{(1-2d)} - 1 \right) 24d \tag{10}$$

Formally, this fixes the $g$-factor to 1. But because the population of rest particles is local and not moving, it can be chosen any other density $d_0$ with the only restriction that this density is large enough to activate the rest particles for the highest velocity. Since it will be seen later that also the density $d$ has no influence on our results, both $d$ and $d_0$ can act as parameters for additional physical modelling, e.g. in multi-phase or multi-component flows.
The equilibrium distributions can now be written as

$$N_i^{eq} = d + \frac{\rho}{24} \left[ 2(c_i)_\alpha u_\alpha + 3(c_i)_\alpha (c_i)_\beta u_\alpha u_\beta - u^2 \right] \tag{11a}$$

for the moving particles and

$$N_0^{eq} = d_0 - \frac{\rho u^2}{2} \tag{11b}$$

for the rest particles. The population of the particles $N_0$ relaxes into its equilibrium with the relaxation time of the system, i.e. with $\tau_{rel} = -\frac{1}{\lambda}$. Therefore, the time evolution of the population of rest particles is given by:

$$N_0(\vec{r}_*, t_* + 1) = N_0(\vec{r}_*, t_*) + \lambda (N_0 - N_0^{eq}) \tag{12a}$$

Because the rest particles are activated isotropically into moving particles, the time evolution of the populations of the moving particles has the following form:

$$\begin{aligned} N_i(\vec{r}_* + \vec{c}_i, t_* + 1) = N_i(\vec{r}_*, t_*) + \mathcal{A}_{ij}(N_j - N_j^{eq}) \\ - \frac{\lambda}{24}(N_0 - N_0^{eq}) \end{aligned} \tag{12b}$$

In this paper I use a projection of the FCHC lattice on two dimensions which is described in detail in [8]. In total there are nine different populations for the moving particles, four in the axial directions ($i = 1, 3, 5, 7$) with $\vec{c}_i = \left( \cos \frac{i-1}{4}\pi, \sin \frac{i-1}{4}\pi \right)$ which have a weight of 4 due to the fact that each is the projection of four links from the original space, four in the diagonal direction ($i = 2, 4, 6, 8$) with $\vec{c}_i = \left( \cos \frac{i-1}{4}\pi, \sin \frac{i-1}{4}\pi \right)$, and one perpendicular to the plane, which has also a weight of 4. The tenth population is that of rest particles. Therefore, the total density is given by $\rho = \sum_{i\,even} N_i + 4 \sum_{i\,odd} N_i$. Introducing new populations via $N_i' \stackrel{!}{=} N_i - d$ and $N_0' \stackrel{!}{=} N_0 - d_0$ and remembering that, due to the symmetry of the collision matrix, $\mathcal{A}_{ij}(N_i - N_i^{eq})$ can be replaced by $\mathcal{A}_{ij}(N_i - N_i^{eq,2})$ [8], equation (10$a$) and (10$b$) can be rewritten in a form, which is more convenient for computational purposes:

$$N_0'(\vec{r}_*, t_*) = N_0'(\vec{r}_*, t_* - 1) + \lambda\left(N_0' + \frac{\rho u^2}{2}\right) \tag{13a}$$

$$N_i'(\vec{r}_*, t_*) = N_i'(\vec{r}_* - \vec{c}_i, t_* - 1) + \mathcal{A}_{ij}\left[N_j'(\vec{r}_*, t*-1) - \frac{\rho}{8}(c_i)_\alpha (c_i)_\beta u_\alpha u_\beta + \frac{\rho}{24}u^2\right] - \frac{\lambda}{24}\left(N_0' + \frac{\rho u^2}{2}\right) \tag{13b}$$

Instead of calculating the total density we now compute $\delta\rho = \sum_{i\,even} N_i' + 4 \sum_{i\,odd} N_i'$. The velocity is still given by expression $\vec{u} = \frac{1}{\rho}\sum_i N_i \vec{c}_i$.

The hydrodynamical system is described by the Reynolds number $Re = \frac{l_0 u_0}{\nu}$ with the characteristic velocity $u_0$ and the number of mesh points $l_0$ for the characteristic length. As a consequence, two parameters out of $\nu$, $l_0$, or $u_0$ can be chosen while the third is fixed by the Reynolds number. The used LBM implies some restrictions to the velocity, where an upper limit of $u = 0.2$ (Mach number $M = 0.28$) has been found [9], and to the viscosity where lower limit of $\nu = 0.14$ has been observed [9]. The upper limit for velocity results from the compressibility of the used lattice gas, because the spatial change of the density describes the pressure (see equation (6)). If the velocity $u$ becomes too large the compressibility effect is not negligible any more. The lower limit for the viscosity has its origin in the relaxation behaviour of the system, because $-\lambda = \frac{2}{6\nu+1}$ is the relaxation parameter $\omega$ and tends to 2 if the viscosity is decreased towards zero. If the overrelaxtion ($\omega > 1$) is too large the system becomes unstable. The lower limit for $\nu$ mentioned above was derived from the numerical experience with the LBM without rest particles. The behaviour of the LBM with rest particles is discussed in section III.

From now on I change the notation in the following way: parameters in the LBM are labeled with $_{LB}$ to distinguish them from the values in the 'real world'. I define $\hat{u}$ and $\hat{l}$ via

$$\vec{u} = \vec{u}_{LB} \cdot \hat{u} \qquad \text{and} \qquad l = l_{LB} \cdot \hat{l} \tag{14a}$$

respectively. If the normalized pressure $p = \frac{p'}{\rho}$ is used, where $p'$ is pressure and $\rho$ the density, the normalized pressure can be obtained from the variation $\delta\rho_{LB}$ in the mean density $\overline{\rho_{LB}}$ via:

$$p = \frac{\delta\rho_{LB}}{\rho_{LB}} \cdot \frac{\hat{u}^2}{2} \tag{14b}$$

The relation for a body force $\vec{f}$ is:

$$\vec{f} = \frac{\hat{l}}{\hat{u}^2}\vec{f}_{LB} \tag{14c}$$

In the following I will give a short description how I implement the boundary conditions. I impose the boundaries on half way between two nodes. At the left, right, and bottom boundary there are no-slip conditions and particles are bounced back. This takes place for all Direchlet boundary conditions except that for non-zero velocities one has to add the right momentum to the particles travelling on the diagonal links. Considering the top boundary, the population $N_3$ is copied into $N_7$ without adding an extra momentum. The population $N_2$ is copied into $N_6$ and $\frac{\overline{\rho_{LB}}}{6} u_{LB}^{top}$ is subtracted. $u_{LB}^{top}$ is the (local) velocity at the top given by the boundary condition. A similar process occurs to the population $N_4$, which is converted into $N_8$ and $\frac{\overline{\rho_{LB}}}{6} u_{LB}^{top}$ is added.
In the case of Neumann boundary conditions particles travelling on diagonal links undergo a specular reflection. The required derivative at the boundary $\delta u_{LB}^{top}$ is achieved by adding (or subtracting) the moment loss due to viscous stress $\frac{\overline{\rho_{LB}}}{2} \nu_{LB} \delta u_{LB}^{top} \equiv \frac{\overline{\rho_{LB}}}{2} \nu_{LB} \left. \frac{\partial u_{LB}(x,y)}{\partial y_{LB}} \right|_{top}$ to the reflected particles.
At this point I have to make a comment about the density used in equation (13a) and (13b) and in the update at the boundary. Within the numerical errors I do not observe a difference between using the local density $\rho_{LB}$ and using the mean density $\overline{\rho_{LB}}$ in (13a) and (13b) for the calculations of the test-problem (see next paragraph). This justifies to use the mean value also for the update of the boundary conditions. Using $\overline{\rho_{LB}}$ instead of the local $\rho_{LB}$ for calculating the equilibrium distribution saves computer time in the collision step.
The great advantage of the LBM's is the local nature of the boundary conditions. Though it can be shown numerically that the Neumann boundary condition implemented in the described way is of second order in space, only the nearest neighbours are involved in its calculation.

After all the theoretical considerations one ends up with a very simple algorithm consisting out of three substeps per time-step:

1. The boundary conditions are set according to the previous time step. The populations $N_i'$ are calculated at the boundary.
   This step involves nearest neighbours.

2. At every node new populations are calculated due to the moving of its nearest neighbours.
   This step involves nearest neighbours.

3. At every node the collision term is calculated and the body forces are added.
   This step is completely local.

For all calculations I use the zero-speed initial condition, i.e. $N_0' \equiv N_i' \equiv 0$.

I run the code on a DEC Alpha station for lattices up to $150 \times 150$ meshpoints and on a NEC SX-3-24R for larger lattices. On the DEC about $10\mu s$ are needed per timestep and node, on the NEC about $83ns$. On the NEC I achieve a performance of between 2.0 and 2.4 GFlops, which has to be related to a peak performance of 6.4 GFlops. The program has not been specially adapted for the NEC computer.

## III. TEST-PROBLEM

To test the LBE method with the previous described boundary conditions I choose a benchmark problem originally proposed by Shih, Tan, and Hwang [6]. The geometry of the problem is a square box of $1m \times 1m$ ($x_{len} = 1m$ and $y_{len} = 1m$) with no-slip boundary conditions at the left, right, and bottom boundary and a shear flow at the top with the velocity profile:

$$u_x(x) = 16(x^4 - 2x^3 + x^2) \tag{15}$$

The maximum speed of $1m/s$ is chosen as the characteristic velocity and therefore the Reynolds number becomes:

$$Re = \frac{1}{\nu} \cdot \frac{m^2}{s} \tag{16}$$

Assuming a time-independent solution of the Navier-Stokes equations, which is true for small Reynolds numbers ($Re < 1000$), the steady state velocity $u^\infty = (u_x^\infty, u_y^\infty)$ is set to:

$$\begin{aligned} u_x^\infty &= \phantom{-}8s(x)\,v'(y) & (17a)\\ u_y^\infty &= -8s'(x)\,v(y) & (17b)\\ \text{with}\quad s(x) &= x^4 - 2x^3 + x^2 & \\ \text{and}\quad v(y) &= y^4 - y^2 & \end{aligned}$$

If the pressure is given by:

$$p^\infty = \nu \cdot 8\big(S(x)v'''(y) + s'(x)v'(y)\big) + 64\big(S_2(x)(v(y)v''(y) - (v'(y))^2\big) \tag{18}$$

the solution of the time-independent Navier-Stokes equations results in a body force $\vec{f} = (f_x, f_y)$:

$$\begin{aligned} f_x &= 0 & (19a)\\ f_y &= \nu \cdot 8\big(24S(x) + 2s'(x)v''(y) + s'''(x)v(y)\big) + 64\big(S_2(x)V(y) - v(y)v'(y)S_1(x)\big) & (19b) \end{aligned}$$

The above used functions are defined as follows:

$$\begin{aligned} S(x) &= \int s(x)dx \\ S_1(x) &= s(x)s''(x) - (s'(x))^2 \\ S_2(x) &= \int s(x)s'(x)dx \\ V(y) &= v(y)v'''(y) - v'(y)v''(y) \end{aligned}$$

I start with a detailed study at $Re = 50$ to check the influence of the density $\rho_{LB}$, the (characteristic) velocity $u_{LB}$ and the lattice size on the accuracy of the results.

The difference between the theoretical and the numerical results is characterized by the relative error of the velocity and the pressure

$$\Delta u = \frac{\sqrt{\sum (u_x - u_x^\infty)^2 + (u_y - u_y^\infty)^2}}{\sqrt{\sum (u_x^\infty)^2 + (u_y^\infty)^2}} \qquad (20a)$$

and

$$\Delta p = \frac{\sqrt{\sum (p - p^\infty)^2}}{\sqrt{\sum (p^\infty)^2}} \qquad (20b)$$

respectively. I do not find an influence of the density $\rho_{LB}$ or the relation $d_0/d$ on the results. The errors for both velocity and pressure become smaller with increasing lattice size and decreasing velocity as it is shown in Fig.1. For the fixed velocity $u_{LB} = 0.02$ the error decreases approximately linear with the mesh size, for higher velocities the slope is even less. A linear behaviour was also observed for calculations with the LBGK method ($u_{LB} = 0.1$) [10]. For a fixed mesh size the axis of velocity can be reinterpretated as the timestep $\Delta t$ in the 'real world', because the timestep is given by $\Delta t = u_{LB}/l_y$ where $l_y$ is the number of mesh points in the vertical.
The required computer time increases linearly with decreasing velocity $u_{LB}$, but it is proportional to the third power of the lattice size. Therefore, one would prefer to reduce the velocity rather than increasing the lattice size but here one is limited to $(u_{LB})_{min} = \frac{Re \cdot \nu_{min}}{l_y}$. From Fig.1 one can deduce that there exist a minimum for the error of the velocity at a certain $(u_{LB})_{min}$. Numerically it is found that this value is smaller the larger the lattice is. The behaviour of the pressure is slightly different und the minimum lies at higher values of $u_{LB}$. Going beyond this value, irregularities in the pressure field grow in the left upper corner. As an example the results of a calculation for $\nu_{LB} = 2.5 \cdot 10^{-3}$ and $l_y = 50$ for $Re = 100$ are shown in Fig.4a. Though the error of the pressure field is rather small (see Table I), some abnormal behaviour of the pressure is observed. The velocity field instead exhibits no unusual featurs compared to those of Fig.3 and it is still very close to the analytical solution. Using the original LBM without rest particles the origin of the vortex is shifted (see Fig.4b) and the calculation loose its validity.[3] For larger values of the viscosity I do not find a difference between the two kinds of methods within the numerical errors. This means that the reservoir of rest particles stabilizes the system for short relaxation times. With the LBGK method including rest particles Hou et.al. reach a lower limit of the viscosity of about $\nu_{LB} = 2.56 \cdot 10^{-3}$ [10]. For their calculations with a fixed top velocity of $u_{LB} = 0.1$ they used a $256 \times 256$ lattice for all Reynolds numbers so that for e.g. $Re = 1000$ the viscosity is $\nu = 0.0256$. There is no possibility to compare their results for the pressure with other numerical calculations nor did they publish a comparison of the pressure field with different values for the viscosity by changing the lattice size.

[3] In this model the pressure cannot be calculated directly via equation (14b).

For the Reynolds number $Re = 100$ I perform three calculations with lattice sizes $50 \times 50$, $100 \times 100$, and $150 \times 150$. The behaviour of the errors in time is shown in Fig.2a. Since the momentum is mainly transported by diffusion, the error of the velocity decreases exponentially in time and the exponent is proportional to the viscosity. Therefore, one expects a similar time-behaviour of the error for $Re = 10$ on a time scale of one magnitude smaller. Indeed, a calculation for $Re$=10 with a lattice size of $50 \times 50$ and $u_{LB}$=0.04 shows this behaviour of the error (see Fig.2b). The oscillatory behaviour of the errors can be reduced by decreasing both the velocity $u_{LB}$ and the viscosity $\nu_{LB}$ keeping the lattice size fixed. The reason for the oscillitory behaviour might be the long relaxation time (high viscosity) and the large timestep (high velocity), which may result in a problem to relax the system into its local equilibrium in every timestep. In Fig.2b the errors of my calculations are compared with those of a finite element calculation with a fractional step $\Theta$ scheme where the lattice size was $64 \times 64$ [11].[4]

I go beyond the Reynolds number of 100 and perform calculations for $Re = 500$, 1000, 2000, and 5000. For $Re = 5000$ computation becomes extremely time consuming because of the large lattice size. I try to reduce the lattice size to $1000 \times 1000$ by choosing three different parameter configurations: a) $u_{LB} = 0.1$ and $\nu_{LB} = 0.02$, b) $u_{LB} = 0.05$ and $\nu_{LB} = 0.01$, and c) $u_{LB} = 0.025$ and $\nu_{LB} = 0.005$. In the first two cases a blow up of the density in the upper right corner takes place after some thousand iterations. While in these both cases the timestep in real units is larger than for the calculation for the $2000 \times 2000$ lattice (a) $\Delta t = 1 \cdot 10^{-4}$, b) $\Delta t = 5 \cdot 10^{-5}$) it is the same in the third case ($\Delta t = 2.5 \cdot 10^{-5} s$). In the third case due to the small viscosity unphysical pressure fluctuations arise on the right-hand side and lead to problems calculating the collision term. From the calculations with the $2000 \times 2000$ mesh I observe a deminishing of the vorticity with time but it is not yet clear if the system will converge towards a steady-state like solution with only small perturbations in time, or if the center of the main vortex will oscillate and some extra vorticities will remain. I have to stress that this case is beyond the validity of the analytical solution and the behaviour of the system is not yet known.

Instead of applying the Dirichlet boundary condition at the top one can force the flow by the Neumann bound-

[4] Only the error for the velocity was available.

ary condition:

$$\left.\frac{\partial u_x(x,y)}{\partial y}\right|_{y=1} = 80(x^4 - 2x^3 + x^2) \qquad (21)$$

I test this boundary condition for Reynolds numbers $Re = 50$, $Re = 100$ and $Re = 1000$. The results are listed in Table II, which shows that the Dirichlet and the Neumann boundary condition give results of nearly the same accuracy. The convergence rate for the errors of the calculation for $Re = 100$ can be seen in Fig.2a. It is of the same order as for the Dirichlet boundary condition. The fact that the LBM yield the result regardless of the kind of boundary condition is very striking in the sense that the numerical effort for both boundary conditions is the same while for the classical methods it is larger for the Neumann boundary condition than for the Dirichlet. The very good agreement of our numerical results with the analytical solution encourage me to use the Neumann boundary condition later for the extended cavity problem.

## IV. EXTENDED DRIVEN CAVITY PROBLEM

In the previous section I have verified that the LBM with enhanced collisions and rest particles reproduces the flow in a lid-driven cavity very nicely. Now I turn to study the flow in a cavity without applying the body force and extend our studies to aspect ratios $Ar := \frac{x_{len}}{y_{len}}$ different from 1. I still apply the same Dirichlet, or in a second step the same Neumann boundary condition at the top of the cavity as in the test-problem. The two different boundary conditions have no longer the same meaning because velocity field is now unknown and not given by equation (17). I have to be a little bit more specific about the geometry of the system and the boundary conditions: the y-axis is fixed to length 1 for all calculations ($y_{len} = 1m$). Defining a new variable $\tilde{x} = Ar^{-1} \cdot x$, with $0 \leq \tilde{x} \leq 1$, the Dirichlet boundary condition is written as

$$u_x(x,1) = 16(\tilde{x}^4 - 2\tilde{x}^3 + \tilde{x}^2) \qquad (22)$$

and the Neumann boundary condition as

$$\left.\frac{\partial u_x(x,y)}{\partial y}\right|_{y=1} = 80(\tilde{x}^4 - 2\tilde{x}^3 + \tilde{x}^2) \qquad (23).$$

Therefore, in the case of the Dirichlet boundary condition the maximum velocity at the top is still $1m/s$ (and the average velocity $0.533m/s$) but the velocity gradient in the x-direction decreases as the aspect ratio increases:

$$\frac{\partial u_x(x,1)}{\partial x} = 32Ar^{-1}(2\tilde{x}^3 - 3\tilde{x}^2 + \tilde{x}) \qquad (24)$$

In other words the velocity field becomes smoother for larger aspect ratios. The same is true for the Neumann boundary condition.

I perform runs for two different values of the viscosity ($\nu = 0.02m^2s^{-1}$ and $\nu = 0.001m^{-2}s^{-1}$, which correspond for the Dirichlet boundary condition to the Reynolds numbers $Re = 50$ and $Re = 1000$) and for four different aspect ratios ($Ar$=0.2, 1, 10 and 50). I also perform calculations with the Dirichlet boundary condition at those Reynolds numbers, which I observe with the Neumann boundary condition and $\nu = 0.001m^{-2}s^{-1}$ at the particular aspect ratio. Table III lists up all the calculations with the used parameters and the origin of the vortices.

I start discussing the results for the small Reynolds number. Without the body force the velocity field is no longer symmetric according to the plane at $x = 0.5m$ but the origin of the vortex is now at $\vec{r} = (0.56, 0.76)m$ instead of $\vec{r} = (0.5, 0.707)m$ for the test-problem. With increasing aspect ratio the vortex is stretched and it is not very helpful to define an origin of a vortex in this case. For $Ar = 50$ the velocity profile for $u_x$ is parabolic except in the vicinity of the left and right wall. The inflection line between forward and back flow lies very near the theoretical value ($y = 2/3\, y_{len}$) for an infinitily long cavity with a constant flow (constant top velocity). In Fig.5 the velocity profile of $u_x$ is shown at vertical cuts through the cavity. The deviation of the velocity at the top and the resulting fluid exchange in the vertical is too small to disturb this flow structure.

A different behaviour of the flow with increasing aspect ratio is observed for a Reynolds number of $Re = 1000$. For the aspect ratio $Ar = 1$ the origin of the vorticity is observed near the centre of the cavity ($\vec{r} = (0.54, 0.573)m$) as it is shown in Fig.6. The main driving force of the system is now the convective momentum transfer. The deep pressure at the centre balances the Centrifugal force of the strong circulation. Pressure and velocity field are quite similar to those of the driven cavity problem with a fixed top velocity [10].

The increase of the aspect ratio to $Ar = 10$ does not lead to just a stretching of the vortex but a vortex is created near the right-hand side of the cavity (its origin is at $\vec{r} = (8.84, 0.522)m$ ). The vorticity of this vortex is much stronger than those of the vortex present in the case of $Ar = 1$, because the main part of the energy, which is put into system by the forced flow at the top, goes into this vortex. The energy of the vortex is so high that the fluid leaving the vortex has enough kinetic energy to run against the pressure gradient of the flow (see Fig.7). The situation changes totally if the aspect ratio is further increased to a value of $Ar = 50$. The vortex on the right-hand side has now disappeared (see Fig.8). I presume that the deviation of the velocity at the top is too smooth to cause a vortex structure. Nevertheless, the system is unstable in the sense that pressure waves are observed travelling in the horizontal direction. The origin of these waves is not yet clear and is a topic of recent research. Another part of the future studies is the question, at which aspect ratio does the transition from a structure with a vortex to a vortex-free structure occur. The results of these investigations will be published elsewhere.

I now turn to present the results for the Neumann boundary condition at the top and the low viscosity. For an aspect ratio of $Ar = 1$ a maximum top speed of $u_{max} = 0.295 m/s$ is found, which means that the Reynolds number is $Re = 295$. Computation of the pressure and velocity field with the Dirichlet boundary condition at this Reynolds number shows that the type of the boundary condition has nearly no influence on the resulting pressure and velocity (see also Table III). The small shift of the maximum top-speed towards the right-hand side in the case of the Neumann boundary condition has no effect on the structure of the vortex. The situation changes if the aspect ratio is $Ar = 10$. In Fig.10 two vortices are found on the right-hand side, a large one similar to those for the Dirichlet boundary condition with its origin at $\vec{r} = (9.16, 0.54)m$ and a small one at the upper left part of the large one with its origin at $\vec{r} = (7.59, 0.63)m$. In accordence with the difference in the velocity field the pressure field exhibits now two high pressure cusps instead one for the Dirichlet boundary condition. In this case the maximum speed is $u_{max} = 0.58 m/s$ corresponding to a Reynolds number of $Re = 580$. For the calculations with the Dirichlet boundary condition no difference in the flow and pressure field on principal can be found between $Re = 1000$ and $Re = 580$ (Fig.11). In both cases a high pressure zone is build up at the right-hand side excluding the main flux from this area. This area is missing for the Neumann boundary condition.
One possibilty to increase the Reynolds number is increasing the deviation of $u_x$ at the top, given by equation (24), by a certain factor. If a factor 2 is chosen a Reynolds number $Re = 830$ is observed with a flow and pressure pattern shown in Fig.12. The second high pressure cusp has disappeared nearly completely and the smaller vortex is shifted towards the inflection line between the flows of opposite direction. In Fig.13 the dependency of the top-velocity on the horizontal position is presented for different Reynolds numbers. For high Reynolds numbers the dependency is quite different from that for the Dirichlet boundary condition while for the low Reynolds number only the center is shifted.
If the aspect ratio is increased to $Ar = 50$, the velocity field is qualitively the same as for the Dirichlet boundary condition. The Reynolds number is $Re = 880$. This case exhibits also pressure fluctuations.

Neumann boundary conditions are of great interest in the case where the fluid in the cavity is coupled to a moving gas at its top. For example, in shallow water simulations the derivative at the top is proportional to the velocity of the wind. In Fig.13 the Neumann boundary condition on top of the box is plotted with the velocity and pressure field, and this distribution can be regarded as the velocity of the wind up to a certain factor. In the future the calculations will be extended to three dimen-sions. In principal, there should be no difference in the numerical behaviour with respect to the two-dimensional case because the method, used in this paper, is a projection of the three-dimensional version.

## V. CONCLUSION

In this paper I have shown that the lattice Boltzmann method with enhanced collisions and rest particles is a fast and accurate method for calculating the velocity and the pressure in a driven cavity by comparing our numerical results with the analytical solution of the problem. Using the Dirichlet or the Neumann boundary condition for the forced flow at the top makes no difference according to the convergence of the numerical results. This is a very striking observation because the implementation of the Neumann boundary condition in the LBM is extremely easy and involves only nearest neighbours. It is possible to decrease the viscosity beyond the value $\nu_{LB} = 0.014$, which was found to be the lower limit for the LBM without rest particles. The pressure is more sensitive to relaxation problems than the velocity.
In the second part of this paper I have presented the results for a driven cavity problem with a non-uniform top speed and aspect ratios different from one. For small Reynolds numbers ($Re = 50$) the result of increasing the aspect ratio is a stretching of the original vortex. At higher Reynolds numbers ($Re = 580, 1000$) a strong vortex on the right-hand side of the cavity has been observed for an aspect ratio of $Ar = 10$. At this aspect ratio a significant difference has been found between the Dirichlet and the Neumann boundary condition because a second but smaller vortex does exist if the Neumann boundary condition are applied at the top.

## ACKNOWLEDGEMENTS

I would like to thank G. Zanetti and R. Benzi for many fruitful discussions. I am much indepted for the high amount of computer time on the NEC SX3-42R supercomputer offered by the CSCS in Manno (Switzerland). This work has been carried out with the financial contribution of Sardinia Regional Authorities.

TABLE I. Parameters and errors for the test-problem with the Dirichlet boundary condition at the top. $\Delta t$ is the timestep of the calculation in terms of the 'real world'. The total time indicates the time in which the system has been converged into its steady-state within the numerical errors except for $Re = 5000$ where the calculation is extremely time consuming.
The CPU-time shows the consumed computer time on a DEC-Alpha station ($\alpha$) or a NEC SX-3-42R (SX-3).
Not all calculations for $Re = 50$ are listed.

| $Re$ | $u_{LB}$ | $l_y$ | $\nu_{LB}$ | $\Delta t$[s] | total time [s] | CPU-time [min] | $\Delta u$ | $\Delta p$ |
|---|---|---|---|---|---|---|---|---|
| 10 | 0.04 | 50 | 0.2 | $8 \cdot 10^{-4}$ | 2 | 1 ($\alpha$) | $2.1 \cdot 10^{-3}$ | $5.8 \cdot 10^{-3}$ |
| | 0.004 | 50 | 0.02 | $8 \cdot 10^{-5}$ | 2 | 11 ($\alpha$) | $1.1 \cdot 10^{-3}$ | $6.8 \cdot 10^{-3}$ |
| 50 | 0.02 | 50 | 0.02 | $4 \cdot 10^{-4}$ | 10 | 11 ($\alpha$) | $1.7 \cdot 10^{-3}$ | $3.9 \cdot 10^{-3}$ |
| | 0.01 | 100 | 0.02 | $1 \cdot 10^{-4}$ | 10 | 167 ($\alpha$) | $4.1 \cdot 10^{-4}$ | $8.4 \cdot 10^{-4}$ |
| 100 | 0.04 | 50 | 0.02 | $8 \cdot 10^{-4}$ | 20 | 11 ($\alpha$) | $4.7 \cdot 10^{-3}$ | $5.0 \cdot 10^{-3}$ |
| | 0.005 | 50 | 0.0025 | $1 \cdot 10^{-4}$ | 20 | 84 ($\alpha$) | $2.8 \cdot 10^{-3}$ | $9.7 \cdot 10^{-3}$ |
| | 0.02 | 100 | 0.02 | $2 \cdot 10^{-4}$ | 20 | 167 ($\alpha$) | $1.2 \cdot 10^{-3}$ | $1.3 \cdot 10^{-3}$ |
| | 0.02 | 150 | 0.03 | $1.33 \cdot 10^{-4}$ | 20 | 563 ($\alpha$) | $7.6 \cdot 10^{-4}$ | $8.4 \cdot 10^{-4}$ |
| 500 | 0.1 | 100 | 0.02 | $1 \cdot 10^{-3}$ | 100 | 167 ($\alpha$) | $3.4 \cdot 10^{-2}$ | $3.3 \cdot 10^{-2}$ |
| | 0.04 | 250 | 0.02 | $1.6 \cdot 10^{-4}$ | 100 | 55 (SX-3) | $5.4 \cdot 10^{-3}$ | $5.5 \cdot 10^{-3}$ |
| 1000 | 0.0787 | 254 | 0.02 | $3.098 \cdot 10^{-4}$ | 124 | 42 (SX-3) | $2.8 \cdot 10^{-2}$ | $2.8 \cdot 10^{-2}$ |
| 2000 | 0.08 | 500 | 0.02 | $1.6 \cdot 10^{-4}$ | 80 | 167 (SX-3) | $5.4 \cdot 10^{-2}$ | $5.1 \cdot 10^{-2}$ |
| 5000 | 0.05 | 2000 | 0.02 | $2.5 \cdot 10^{-5}$ | 12.5 | 2770 (SX-3) | $7.7 \cdot 10^{-1}$ | $1.2 \cdot 10^{0}$ |

TABLE II. Parameters and errors for the test-problem with the Neumann boundary condition at the top. $\Delta t$ is the timestep of the calculation in terms of the 'real world'. The total time indicates the time in which the system has been converged into its steady-state within the numerical errors.
The CPU-time shows the consumed computer time on a DEC-Alpha station ($\alpha$) or a NEC SX-3-42R (SX-3).

| $Re$ | $u_{LB}$ | $l_y$ | $\nu_{LB}$ | $\Delta t$[s] | total time [s] | CPU-time [min] | $\Delta u$ | $\Delta p$ |
|---|---|---|---|---|---|---|---|---|
| 50 | 0.02 | 50 | 0.02 | $4 \cdot 10^{-4}$ | 10 | 11 ($\alpha$) | $1.7 \cdot 10^{-3}$ | $2.8 \cdot 10^{-3}$ |
| 100 | 0.04 | 50 | 0.02 | $8 \cdot 10^{-4}$ | 20 | 11 ($\alpha$) | $3.8 \cdot 10^{-3}$ | $3.9 \cdot 10^{-3}$ |
| 1000 | 0.0787 | 254 | 0.02 | $3.098 \cdot 10^{-4}$ | 124 | 42 (SX-3) | $4.5 \cdot 10^{-2}$ | $4.0 \cdot 10^{-2}$ |

TABLE III. Parameters and results for the extended driven cavity problem.
$Re$ is the Reynolds number, $Ar$ the aspect ratio, bc denotes the boundary condition (D: Dirichlet, N: Neumann), $\vec{r}$ are the coordinates of the origin of a vortex, and $u_{max}$ is the maximum speed (allways at $y = 1m$).

| $\nu$ | $Re$ | $Ar$ | bc | mesh size | $u_{LB}$ | $\vec{r}$ of 1.vortex | $\vec{r}$ of 2.vortex | $u_{max}[m/s]$ | at $x[m]$ |
|---|---|---|---|---|---|---|---|---|---|
| 0.02 | 50 | 0.2 | D | 40×200 | 0.01 | (0.103,0.95)$m$ | – | 1.0 | 0.1 |
| 0.02 | 50 | 1.0 | D | 100×100 | 0.01 | (0.56,0.76)$m$ | – | 1.0 | 0.5 |
| 0.02 | 50 | 10 | D | 1000×100 | 0.01 | – | – | 1.0 | 5.0 |
| 0.02 | 50 | 50 | D | 10000×100 | 0.01 | – | – | 1.0 | 25.0 |
| 0.02 | 6.1 | 0.2 | N | 40×200 | 0.0022 | (0.100,0.95)$m$ | – | 0.121 | 0.103 |
| 0.02 | 29 | 1.0 | N | 100×100 | 0.0057 | (0.57,0.78)$m$ | – | 0.573 | 0.555 |
| 0.02 | 57 | 10 | N | 1000×100 | 0.0011 | – | – | 1.14 | 5.91 |
| 0.001 | 1000 | 0.2 | D | 40×200 | 0.1 | (0.120,0.94)$m$ | – | 1.0 | 0.1 |
| 0.001 | 1000 | 1.0 | D | 254×254 | 0.0787 | (0.54,0.57)$m$ | – | 1.0 | 0.5 |
| 0.001 | 1000 | 10 | D | 2540×254 | 0.0787 | (8.84,0.52)$m$ | – | 1.0 | 5.0 |
| 0.001 | 1000 | 50 | D | 25400×254 | 0.0787 | – | – | 1.0 | 25.0 |
| 0.001 | 118 | 0.2 | N | 40×200 | 0.0118 | (0.110,0.95)$m$ | – | 0.118 | 0.108 |
| 0.001 | 295 | 1.0 | N | 254×254 | 0.0232 | (0.60,0.64)$m$ | – | 0.295 | 0.66 |
| 0.001 | 580 | 10 | N | 2540×254 | 0.0457 | (9.16,0.54)$m$ | (7.59,0.63)$m$ | 0.58 | 7.4 |
| 0.001 | 830 | 10 | N | 2540×254 | 0.0457 | (9.35,0.55)$m$ | (7.81,0.77)$m$ | 0.83 | 6.6 |
| 0.001 | 880 | 50 | N | 25400×254 | 0.0693 | – | – | 0.88 | 32.0 |
| 0.001 | 118 | 0.2 | D | 40×200 | 0.0118 | (0.105,0.95)$m$ | – | 0.118 | 0.1 |
| 0.001 | 295 | 1.0 | D | 254×254 | 0.0232 | (0.60,0.64)$m$ | – | 0.295 | 0.5 |
| 0.001 | 580 | 10 | D | 2540×254 | 0.0457 | (8.07,0.55)$m$ | – | 0.58 | 5.0 |

FIG. 1. Relative errors for velocity and pressure according to the definition of equation (20) for $Re = 50$ and different parameter configurations. The axis for the velocity $u_{LB}$ is logarithmic for a better visualisation of the results. The lines on the basic plane mark contours of constant viscosity $\nu_{LB}$. The values vor $\nu_{LB}$ are from left to right: 0.0025, 0.005, 0.01, 0.05, 0.1, and 0.2.

FIG. 2. Relative errors for velocity and pressure according to the definition of equation (20) versus the time in the 'real world' for $Re = 100$ (left) and $Re = 10$ (right). Lines with points in its style belong to the pressure, the others to the velocity.
For $Re = 100$ all results as listed in Table I are plotted except the case with the viscosity $\nu = 0.0025$. Also the result with the Neumann boundary condition is plotted (compare Table II).
For $Re = 10$ the results for $u_{LB} = 0.04$ (solid line for velocity and pointed line for pressure) and $u_{LB} = 0.004$ (dashed line for velocity and dashed-pointed for pressure) are shown.
The line with the stars as markers is the result of a finite element calculation [11].

FIG. 3. Pressure and velocity profile for the test-problem ($Re = 100$).

FIG. 4. Pressure and velocity profile for the test-problem ($Re = 100$). The viscosity is $\nu_{LB} = 0.0025$.

FIG. 5. Velocity profiles for $u_x$ at different vertical cuts through the cavity for $Re = 50$ and $Ar = 50$.

FIG. 6. Pressure and velocity profile for $Re = 1000$, aspect ratio $Ar = 1$ and the Dirichlet boundary condition at the top.

FIG. 7. Pressure and velocity profile for $Re = 1000$, aspect ratio $Ar = 10$, and the Dirichlet boundary condition at the top.
The aspect ratio of the picture does not agree with those of the calculation.

FIG. 8. Velocity profile for $Re = 1000$, aspect ratio $Ar = 50$, and the Dirichlet boundary condition at the top. The aspect ratio of the picture does not agree with those of the calculation.

FIG. 9. Pressure and velocity profile for $Re = 295$, aspect ratio $Ar = 1$, and the Neumann boundary condition at the top.

FIG. 10. Pressure and velocity profile for $Re = 580$, aspect ratio $Ar = 10$, and the Neumann boundary condition at the top. The aspect ratio of the picture does not agree with those of the calculation.

FIG. 11. Pressure and velocity profile for $Re = 580$, aspect ratio $Ar = 10$, and the Dirichlet boundary condition at the top. The maximum speed at the top is reduced to $0.58 m/s$, the viscosity is $\nu = 0.001 m^2/s$.
The aspect ratio of the picture does not agree with those of the calculation.

FIG. 12. Pressure and velocity profile for $Re = 830$, aspect ratio $Ar = 10$, and the Neumann boundary condition at the top. The maximum derivative at the top is increased to $10 s^{-1}$, the viscosity is $\nu = 0.001 m^2/s$.
The aspect ratio of the picture does not agree with those of the calculation.

FIG. 13. Velocity at the top for different Reynolds numbers when the Neumann boundary condition is applied at the top. The aspect ratio is $Ar = 10$, the viscosity $\nu = 0.001 m^2/s$ for the high Reynolds numbers, and $\nu = 0.02 m^2/s$ for the low Reynolds number.